\documentclass[aps,nofootinbib,twocolumn,prd,eqsecnum,showpacs,showkeys,preprintnumbers]{revtex4-1}

\usepackage[caption=false]{subfig}
\usepackage{graphicx}
\usepackage{amsmath}
\usepackage{amsfonts}
\usepackage{amssymb}
\usepackage{color}
\usepackage{bm}
\usepackage{mathrsfs}
\usepackage{epstopdf}
\usepackage{url}
\usepackage{footnote}
\usepackage{textcomp}
\usepackage{dsfont}

\makeatletter
\newcommand*{\rom}[1]{\expandafter\@slowromancap\romannumeral #1@}
\makeatother

\begin{document}

\title{Tensor Perturbations from Brane-World Inflation with Curvature Effects}
\author{Mariam Bouhmadi-L\'{o}pez $^{1,2}$}
\email{mariam.bouhmadi@ehu.es}
\author{Yen-Wei Liu $^{3,4}$}
\email{f97222009@ntu.edu.tw}
\author{Keisuke Izumi $^{4}$}
\email{izumi@phys.ntu.edu.tw}
\author{Pisin Chen $^{3,4,5,6}$}
\email{chen@slac.stanford.edu}
\date{\today}

\affiliation{
${}^1$Department of Theoretical Physics, University of the Basque Country
UPV/EHU, P.O. Box 644, 48080 Bilbao, Spain\\
${}^2$IKERBASQUE, Basque Foundation for Science, 48011, Bilbao, Spain\\
${}^3$Department of Physics, National Taiwan University, Taipei, Taiwan 10617\\
${}^4$Leung Center for Cosmology and Particle Astrophysics, National Taiwan University, Taipei, Taiwan 10617\\
${}^5$Graduate Institute of Astrophysics, National Taiwan University, Taipei, Taiwan 10617\\
${}^6$Kavli Institute for Particle Astrophysics and Cosmology, SLAC National Accelerator Laboratory, Stanford University, Stanford, California 94305, U.S.A.
}

\begin{abstract}
The brane-world scenario provides an intriguing possibility to explore the phenomenological cosmology implied by string/M theory. In this paper, we consider a modified Randall-Sundrum single brane model with two natural generalizations: a Gauss-Bonnet term in the five-dimensional bulk action as well as an induced gravity term in the four-dimensional brane action, which are the leading-order corrections to the Randall-Sundrum model. We study the influence of these combined effects on the evolution of the primordial gravitational waves generated during an extreme slow-roll inflation on the brane. The background, for the early inflationary era, is then modeled through a de Sitter brane embedded within an anti--de Sitter bulk. In this framework, we show that both effects tend to suppress the Randall-Sundrum enhancement of the amplitude of the tensor perturbations at relatively high energies. Moreover, the Gauss-Bonnet effect, relative to standard general relativity, will abruptly enhance the tensor-to-scalar ratio and break the standard consistency relation at high energies, which cannot be evaded by invoking the induced gravity effect, even though the induced gravity strength would mildly counterbalance these significant changes at high energies. We note that the brane-world model with or without the induced gravity effect fulfills the consistency relation. Finally, we discuss some implications of the observational constraints.
\end{abstract}

%%%%%%%%%%%%%%%%%%%%%%%%%%%%%%%%%%%%%%%%%%%%%%%%%%%%%%%%%%%%%%%%%%%%%%%%%%%%%%%%%
\keywords{}
\pacs{}

\maketitle

\section{introduction}

Several brane-world scenarios, inspired by superstring/M theory, have been proposed to resolve various puzzles in particle physics and cosmology, in which our observable four-dimensional (4D) universe may be a hypersurface embedded within a higher-dimensional space-time, usually named a bulk. In this scenario, the Standard Model particles are restricted to the brane, while the gravitons are free to propagate into the whole space (see, for example, the review \cite{Maartens:2010ar}). Among these, the Randall-Sundrum single brane model (RS2 model) \cite{hep-th/9906064} provides a new perspective on the brane-world phenomenology with a noncompact extra dimension, where a single 3-brane is embedded in a five-dimensional (5D) anti--de Sitter (AdS$_5$) bulk. Despite the infinite extra dimension with a continuum of Kaluza-Klein (KK) modes, the 5D graviton zero mode, at sufficiently low energies, is localized on the brane and standard general relativity (GR) is precisely recovered, due to the warped bulk geometry.

In the setup of the RS2 brane-world model, there are two natural generalizations of the gravitational action. The first modification, based on the nature of higher dimensions, is a Gauss-Bonnet (GB) term in the 5D bulk action, leading to the most general second-order derivative field equation of the bulk metric \cite{Lovelock:1971yv}. Moreover, motivated by string/M theory in the low-energy limit, this GB correction corresponds to the leading-order correction in the effective action of heterotic string theory and is free of ghosts \cite{Zwiebach:1985uq,Zumino:1985dp}. Furthermore, the GB term is also essential in the Chern-Simons gauge theory of gravity in odd dimensions \cite{Witten:1988hc,Chamseddine:1989nu,Zanelli:2005sa}. In addition, quantum corrections generated via quantum loops of matter fields on the brane coupled to the bulk gravitons will induce a second modification to the RS2 model, the leading correction of which is
described through an induced gravity (IG) term, i.e., a 4D Ricci scalar, in the brane action \cite{Sakharov:1967pk,Collins:2000yb,Dvali:2000hr,Dvali:2000xg,Shtanov:2000vr,Deffayet:2000uy}.

Since both effects are the leading corrections to the RS2 brane-world model, we investigate here a generalized RS2 model taking into account both the GB and IG curvature effects. In particular, we will look into the early inflationary scenario and study the evolution of the gravitational waves generated during an extreme slow-roll inflation on the brane. In this framework, the background of this scenario is modeled approximately by a de Sitter brane embedded in an AdS$_5$ bulk, on which the primordial gravitational waves will be affected in particular by GB and IG corrections. Cosmological perturbations in a quasi--de Sitter inflation have been previously studied in the RS2 model \cite{Langlois:2000ns,hep-ph/9912464}, and in the generalized RS2 model with just the GB effect \cite{Dufaux:2004qs} or IG effect alone \cite{BouhmadiLopez:2004ax}. In addition, the scalar perturbations in the generalized RS2 model with both the GB and the IG curvature effects have recently been investigated as well in Ref.~\cite{BouhmadiLopez:2012uf}.

Taking both GB and IG curvature effects into consideration, we compare their effects on the behavior of the primordial gravitational waves with those generated in the pure RS2 model. The mass spectrum of KK modes derived in this scenario for the normal branch [see Eq.(\ref{Friedmann}) for $\epsilon=-1$] is exactly the same as that in the RS2 model. In addition, we show that both GB and IG corrections tend to decrease the RS2 enhancement of the amplitude of the tensor perturbations; in fact, the amplitude will be strongly suppressed by the GB effect at relatively high energies. A similar behavior has been obtained for the scalar perturbations in the same system \cite{BouhmadiLopez:2012uf}. Furthermore, the GB effect, as compared with the standard 4D general relativity results, will drastically raise the tensor-to-scalar ratio in the relatively high-energy regime and moderately break the standard consistency relation, which cannot be avoided by incorporating the IG effect. We note that the RS2 model with or without the IG effect fulfills the consistency relation. However, these significant changes caused by the GB effect, will be only mildly counterbalanced by the IG effect at high energies. Finally, we will discuss some implications of the observational constraints on the primordial gravitational waves in this model.

The outline of this paper is the following. In Sec.~\ref{model}, we consider a generalized RS2 model with GB and IG curvature effects, and we review the background solutions of this system. In Sec.~\ref{bulkpert}, we calculate the bulk metric perturbations, where we focus on the three-dimensional (3D) tensor modes on a de Sitter brane embedded within an AdS$_5$ bulk. In Sec.~\ref{KK}, we analyze the spectrum of KK modes. In Sec.~\ref{priGW}, we study the effects from both GB and IG curvature terms on the amplitude of the primordial gravitational waves, and we discuss the tensor-to-scalar ratio in this scenario. In addition, we discuss the observational constraints on this model. Finally, we present our conclusion in Sec.~\ref{conclusion} and include an appendix where we describe the 4D effective action used to obtain the amplitude of the gravitational waves in this model.

\section{The model}\label{model}

We consider a 5D brane-world model with a single brane embedded in a 5D bulk. The bulk action contains a GB term in addition to the usual Hilbert-Einstein term, while the brane action is described by an IG term, a brane tension, and a matter Lagrangian. The gravitational action for this system is given by%\footnote{For simplicity, we have omitted the York-Gibbons-Hawking surface term \cite{York:1972sj,Gibbons:1976ue}, which is needed for the variation principle to be well-defined on the brane.}
\begin{align}
S=&\frac1{2\kappa_5^2}\int_Md^5x\sqrt{-\,^{(5)\!}g}\left[\mathcal{R}-2\Lambda_5+\alpha\left(\mathcal{R}^2-4\mathcal{R}_{ab}
  \mathcal{R}^{ab}+\right.\right.\notag \\
  &\left.\left.\mathcal{R}_{abcd}\mathcal{R}^{abcd}\right)\right]-\frac1{\kappa_5^2}\int_{\Sigma_{\pm}}d^4x\sqrt{-g}\left[K+2\alpha\left(J-\right.\right.\notag\\
  &\left.\left.2G^{\mu\nu}K_{\mu\nu}\right)\right]+\int_{\Sigma}d^4x\sqrt{-g}\left[\frac{\gamma}{2\kappa^2_4}R-\lambda+\mathscr{L}_m\right]\label{action},
\end{align}
where the 5D manifold is split into two parts by a brane hypersurface $\Sigma$, and the two sides of the brane are denoted by $\Sigma_{\pm}$; $g_{ab}=\,^{(5)\!}g_{ab}-n_an_b$ is the induced metric on the brane, in which $n^a$ is the unit normal vector to the brane; the Latin indices $a,b,c,\ldots,$ run from 0 to 4, while the Greek indices $\mu,\nu,\ldots,$ run from 0 to 3; $\mathcal{R}$ is the 5D Ricci scalar, and $R$ in the brane action is the Ricci scalar of the induced metric; $\kappa_5^2$ is the bulk gravitational constant; $\Lambda_5\,(<0)$ is the bulk cosmological constant; and $\lambda$ is the brane tension. The GB parameter $\alpha\,(\geq0)$ has the dimension of length square; while the strength of the IG term is characterized by a dimensionless parameter $\gamma$. In addition, the second term in Eq.(\ref{action}) corresponds to the generalized York-Gibbons-Hawking surface term \cite{York:1972sj,Gibbons:1976ue,Myers:1987yn,Davis:2002gn}, where $K_{\mu\nu}$ is the extrinsic curvature, $G_{\mu\nu}$ the Einstein tensor of the induced metric, and $J$ the trace of
\begin{align}
J_{\mu\nu}=&\frac13\left(2KK_{\mu\sigma}K^{\sigma}\,_{\nu}+K_{\rho\sigma}K^{\rho\sigma}K_{\mu\nu}-2K_{\mu\rho}K^{\rho\sigma}
K_{\sigma\nu}\right.\notag\\
&\left.-K^2K_{\mu\nu}\right).
\end{align}

Then the 5D field equation is obtained by varying the bulk action in Eq.(\ref{action}),
\begin{equation}
\mathcal{G}_{ab}+\Lambda_5 \,^{(5)\!}g_{ab}-\frac{\alpha}2 \mathcal{H}_{ab}=0,\label{fieldeq}
\end{equation}
where $\mathcal{G}_{ab}$ is the 5D Einstein tensor and the quadratic curvature correction $\mathcal{H}_{ab}$ reads
\begin{align}
\mathcal{H}_{ab}=&\left(\mathcal{R}^2-4\mathcal{R}_{cd}\mathcal{R}^{cd}+\mathcal{R}_{cdef}
\mathcal{R}^{cdef}\right)\,^{(5)\!}g_{ab}\notag\\
&-4\left(\mathcal{R}\mathcal{R}_{ab}-2\mathcal{R}_{ac}\mathcal{R}_{b}\,^{c}-2\mathcal{R}_{acbd}\mathcal{R}^{cd}\right.\notag\\
&\left.+\mathcal{R}_{acde}\mathcal{R}_b\,^{cde}\right).
\end{align}

Under the assumption of $\mathds{Z}_2$ symmetry across the brane, the junction condition, relating the discontinuity of the geometry at the brane to its matter content, is given by \cite{Maeda:2003vq,Davis:2002gn,Charmousis:2002rc,Maeda:2007cb},
\begin{align}
&2\left(K_{\mu\nu}-Kg_{\mu\nu}\right)+4\alpha\left(3J_{\mu\nu}-Jg_{\mu\nu}+2P_{\mu\rho\sigma\nu}K^{\rho\sigma}\right)\notag\\
&=-\kappa_5^2 \left(T_{\mu\nu}-\lambda g_{\mu\nu}-\frac{\gamma}{\kappa_4^2}G_{\mu\nu}\right),\label{jc}
\end{align}
where $T_{\mu\nu}$ is the energy-momentum tensor of the matter content on the brane, and
\begin{align}
P_{\mu\nu\rho\sigma}=R_{\mu\nu\rho\sigma}+2R_{\nu[\rho}g_{\sigma]\mu}-2R_{\mu[\rho}g_{\sigma]\nu}+Rg_{\mu[\rho}g_{\sigma]\nu}\,\,.
\end{align}

Since we are mainly interested in the extreme slow-roll inflationary scenario, we study a de Sitter brane embedded in an AdS$_5$ bulk. This solution fully describes the extreme slow-roll regime of the brane where the kinetic energy density of the inflaton confined on the brane can be neglected as compared with the inflaton potential. For an AdS$_5$ bulk, the 5D curvature tensors satisfy
\begin{align}
\mathcal{R}_{abcd}=&-\mu^2\left[^{(5)\!}g_{ac}\,^{(5)\!}g_{bd}-\,^{(5)\!}g_{ad}\,^{(5)\!}g_{bc}\right],\\
\mathcal{H}_{ab}=&24\mu^4\,^{(5)\!}g_{ab}\,,
\end{align}
where $\mu$ is the energy scale associated with the AdS$_5$ length $\ell\equiv1/\mu$. Moreover, the relation (\ref{fieldeq}) implies $\Lambda_5=-6\mu^2+12\alpha\mu^4$; therefore, the energy scale $\mu$ has the solutions
\begin{equation}
\mu^2=\frac1{4\alpha}\left[1\pm\sqrt{1+\frac43\alpha\Lambda_5}\right].\label{mu}
\end{equation}
However, from now on we disregard the $+$ branch on the previous equation as it is unstable \cite{Boulware:1985wk,Myers:1988ze,Cai:2001dz,Charmousis:2008ce}. On the other hand, Eq.(\ref{mu}) with the $-$ sign is interesting because it reduces to the RS2 relation, i.e., $\mu^2=-\Lambda_5/6$, in the absence of a GB correction $\alpha\rightarrow0$. This feature is important as we want to compare our results with the RS2 setup. The energy scale $\mu^2$ is therefore bounded as\footnote{Notice that we have excluded the limiting case where $\mu^2=1/4\alpha$ corresponding to the Chern-Simons gravity, because in that case a homogenous and isotropic brane cannot be embedded in the bulk \cite{Charmousis:2002rc}. Besides, we assume a nonvanishing bulk cosmological constant $\Lambda_5$.} $0<\mu^2<1/4\alpha$.

The background metric can be given in terms of the Gaussian normal coordinate,
\begin{equation}
ds^2=n^2(y)\left[-dt^2+e^{2Ht}d\vec{x}^2\right]+dy^2.
\end{equation}
The brane is located at $y=0$, and the warp factor $n(y)$ is given by
\begin{equation}
n(y)=\frac H{\mu} \sinh{\left[\mu\left(y_*+\epsilon|y|\right)\right]},\label{warp}
\end{equation}
with
\begin{equation}
y_*=\frac1{\mu}\mathrm{arcsinh}{\left(\frac{\mu}H\right)},
\end{equation}
where $\mathds{Z}_2$ symmetry has been imposed, and the normalization of the warp factor is chosen such that $n(0)=1$. This metric form is then a solution of the field equation (\ref{fieldeq}) as well as the junction condition (\ref{jc}) with a constant energy density $\rho\,(>0)$ in $T_{\mu\nu}$ and a constant Hubble parameter $H$.

Different signs $\epsilon=\pm1$ in the warp factor (\ref{warp}) are related to the different effective Friedmann equations on the brane derived through the junction condition (\ref{jc}),
\begin{align}
&\left[1+\frac83\alpha\left(H^2-\frac{\mu^2}2\right)\right]\sqrt{H^2+\mu^2}\notag\\
&=-\epsilon \,r_c\left[\frac{\kappa^2_4}{3}\left(\rho+\lambda\right)-\gamma H^2\right]\label{Friedmann},
\end{align}
where the crossover scale $r_c$ is defined by $r_c\equiv\kappa_5^2/2\kappa_4^2$, and the energy density $\rho$ is taken to be a constant during the quasi--de Sitter inflationary era. In general, there are three evolutionary solutions encoded in the cubic Friedmann equation (\ref{Friedmann}), whose exact solutions have been analyzed in a different context in Refs.~\cite{Kofinas:2003rz,Brown:2006mh,BouhmadiLopez:2008nf,BouhmadiLopez:2009jk,BouhmadiLopez:2011xi,Belkacemi:2011zk} (see also\footnote{We leave the induced gravity parameter $\gamma$ arbitrary, and we do not set to unity as was done in Refs.~\cite{Kofinas:2003rz,Brown:2006mh}.} Ref.~\cite{BouhmadiLopez:2012uf}). In particular, the ``self-accelerating branch'' [$\epsilon=+1$ in Eq.(\ref{Friedmann})] contains the Dvali-Gabadadze-Porrati (DGP) self-accelerating solution modified by the GB effect \cite{Dvali:2000hr,Dvali:2000xg,Deffayet:2000uy,Shtanov:2000vr}, while the ``normal branch'' [$\epsilon=-1$ in Eq.(\ref{Friedmann})] recovers standard GR in the low energy limit and reduces to the RS2 model when both GB and IG effects vanish.

The warp factor $n(y)$, in the self-accelerating branch ($\epsilon=+1$), has its minimum at the brane, and the extradimensional coordinate $y$ could be any real number: $0\leq|y|$. Conversely, in the normal branch ($\epsilon=-1$) the maximum warp factor is reached at the brane, which is similar to the RS2 model. Besides, the coordinate $y$ is bounded as $|y|< y_*$, where $n(y_*)=0$ corresponds to the location of the Cauchy horizon.

Before analyzing the tensor perturbations, we introduce the conformal bulk coordinate $z$ for later convenience: $dz=dy/n(y)$. The bulk metric becomes
\begin{equation}
ds^2=\tilde{n}^2(z)\left[-dt^2+dz^2+e^{2Ht}d\vec{x}^2\right],
\end{equation}
where the conformal bulk coordinate $z$ can be written as a function of $y$,
\begin{equation}
z=-\mathrm{sgn}(y)\frac{\epsilon}H\ln{\coth{\left[\frac12\mu(y_*+\epsilon|y|)\right]}},\label{z}
\end{equation}
and the warp factor $\tilde{n}(z)$ is obtained as
\begin{equation}
\tilde{n}(z)=\frac H{\mu\sinh{H|z|}}.\label{nz}
\end{equation}
In terms of the conformal bulk coordinate, $z$, the bulk space-time in the self-accelerating branch is bounded: $0<|z|\leq z_b$, where $z_b$ is defined by
\begin{equation}
z_b\equiv|z(0)|=\frac1H\ln{\coth{\left(\frac{\mu y_*}2\right)}}.
\end{equation}
Moreover, when approaching the location of the brane as $y\rightarrow0^+$, the conformal bulk coordinate $z\rightarrow -z_b$. On the other hand, the normal branch is mapped into a different extradimensional region $z_b\leq|z|$, and the Cauchy horizons are therefore located at infinity. In addition, as $y\rightarrow0^+$, the coordinate $z\rightarrow z_b$ [see Eq.(\ref{z})].

\section{bulk metric perturbations}\label{bulkpert}

Here we will consider the tensor perturbations and focus on the 3D tensor modes over a de Sitter brane in an AdS$_5$ bulk. The perturbed metric can be written in the form
\begin{equation}
ds^2=n^2(y)\left[-dt^2+e^{2Ht}\left(\delta_{ij}+h_{ij}\right)dx^idx^j\right]+dy^2,\label{perturbedg}
\end{equation}
where the 3D tensor $h_{ij}(t,\vec{x},y)$ is transverse and traceless, and the brane is still located at $y=0$. With this perturbed metric, we obtain the perturbed curvature correction $\delta\mathcal{H}^i\,_j=8\mu^2\delta\mathcal{G}^i\,_j$; thus the perturbed bulk field equation is
\begin{equation}
\left(1-\beta\right)\delta\mathcal{G}^i\,_j=0\Rightarrow\delta\mathcal{G}^i\,_j=0, \label{pfieldq}
\end{equation}
where $\beta$ is a dimensionless parameter defined as $\beta\equiv4\alpha\mu^2$. The resulting field equation (\ref{pfieldq}) implies that the bulk solutions are the same as in the RS2 model \cite{Langlois:2000ns,Bridgman:2001mc}. Although the bulk equation of motion (\ref{pfieldq}) is unchanged compared with the RS2 case at the brane, both GB and IG effects will change the perturbed junction condition, which will further modify the normalization of the modes as shown below [see Eq.(\ref{sp})].

The perturbed junction condition, obtained by using Eqs.(\ref{jc}) and (\ref{perturbedg}), gives
\begin{equation}
\delta K^i\,_j|_{0^+}=\frac1{1-\beta}\left[\gamma r_c-\epsilon\,\beta\mu^{-1}\sqrt{1+x^2}\right]\delta G^i\,_j|_{0^+},\label{perturbedjc}
\end{equation}
where the dimensionless parameter $x$ is defined as $x\equiv H/\mu$, and where we have neglected the anisotropic stress of the matter on the brane. This is a natural assumption as the brane expansion is simply driven by an inflaton and the brane tension $\lambda$. We will later impose a fine-tuning of $\lambda$ on the normal branch \textit{\`{a} la} the RS2 model to see the effects of GB and IG curvature terms on that model. In the limit when $\beta\rightarrow0$ and $\gamma\rightarrow0$, the boundary condition (\ref{perturbedjc}) recovers the case in the RS2 model \cite{Langlois:2000ns}.

Combining the perturbed bulk equation (\ref{pfieldq}) and the junction condition (\ref{perturbedjc}), we obtain the full perturbed field equation of this system, which reads
\begin{align}
&\left[\partial^2_t+3H\partial_t-e^{-2Ht}\nabla^2-\left(n^2\partial^2_y+4nn'\partial_y\right)\right]h_{ij}\notag\\
&=-\frac2{1-\beta}\left[\gamma r_c-\epsilon\,\beta\mu^{-1}\sqrt{1+x^2}\right]\notag\\
&\quad\times\left(\partial^2_t+3H\partial_t-e^{-2Ht}\nabla^2\right)h_{ij}\,\delta\left(y\right)\,,\label{perteq}
\end{align}
where the prime denotes a derivative with respect to $y$. This field equation (\ref{perteq}) can then be separated into an eigensystem problem by decomposing 3D tensor perturbations $h_{ij}$ into KK modes,
\begin{equation}
h_{ij}(t,\vec{x},y)=\int{dm h^{(m)}_{ij}(t,\vec{x})\,\mathcal{E}_m(y)},\label{decomposition}
\end{equation}
where, as usual, ``$\int$\,'' stands for a summation over the discrete modes and an integration over the continuous modes, and $h^{(m)}_{ij}$ and $\mathcal{E}_m$ satisfy
\begin{align}
&\left[\partial^2_t+3H\partial_t-e^{-2Ht}\nabla^2+m^2\right]h^{(m)}_{ij}=0,\\
&\left(n^4\mathcal{E}'_m\right)^{\prime}+m^2n^2\mathcal{E}_m+\frac{2m^2}{1-\beta}\times\notag\\
&\left[\gamma r_c-\epsilon\,\beta\mu^{-1}\sqrt{1+x^2}\right]\mathcal{E}_m \,\delta(y)=0.\label{Em}
\end{align}
In addition, the boundary condition for $\mathcal{E}_{m}$ at the brane can be derived directly from Eq.(\ref{perturbedjc}), which is included in Eq.(\ref{Em}) as well,
\begin{equation}
\mathcal{E}'_m\left(0^+\right)=-\frac{m^2}{1-\beta}\left[\gamma r_c-\epsilon\,\beta\mu^{-1}\sqrt{1+x^2}\right]\mathcal{E}_m\left(0^+\right).\label{bcEm}
\end{equation}
As a result, GB and IG effects give rise to a mass dependence on the boundary condition for $\mathcal{E}_{m}$, and only the massless mode, with $m=0$, has the same boundary condition as in the RS2 model \cite{Langlois:2000ns}.

Furthermore, it can be shown that using Eq.(\ref{Em}), these eigenmodes $\mathcal{E}_{m}$ are orthonormal with respect to the following scalar product:
\begin{align}
\left(\mathcal{E}_{m},\mathcal{E}_{\tilde{m}}\right)=&\left(1-\beta\right)\int^{u_{\epsilon}}_{-u_{\epsilon}}dyn^2\mathcal{E}_{m}
\mathcal{E}_{\tilde{m}}\notag\\
&+2\left[\gamma r_c-\epsilon\,\beta\mu^{-1}\sqrt{1+x^2}\right]\mathcal{E}_m(0)\mathcal{E}_{\tilde{m}}(0)\notag\\
=&\,\delta\left(m,\tilde{m}\right)\label{sp},
\end{align}
where $u_+=\infty$ and $u_-=y_*$. The delta function $\delta\left(m,\tilde{m}\right)$ is a Kronecker delta for discrete modes and a Dirac delta function for a continuous spectrum. Besides, with the scalar product chosen in Eq.(\ref{sp}), the 4D effective action for the metric perturbations is given by \cite{Dufaux:2004qs}
\begin{align}
S_{\mathrm{eff}}^{(2)}=&\frac1{8\kappa^2_5}\int dm \,\mathcal{E}^{-2}_m(0)\int dx^4e^{3Ht}\left[\dot{\bar{h}}^{(m)ij}\,\dot{\bar{h}}^{(m)}_{ij}\notag\right.\\
&\left.-e^{-2Ht}\partial^k\bar{h}^{(m)ij}\,\partial_k\bar{h}^{(m)}_{ij}-m^2\bar{h}^{(m)ij}\,\bar{h}^{(m)}_{ij}\right],\label{effaction}
\end{align}
where the dot denotes a derivative with respect to time $t$, and the tensor modes $\bar{h}^{(m)}_{ij}$ are defined as $\bar{h}^{(m)}_{ij}=h^{(m)}_{ij}\mathcal{E}_m(0)$, corresponding to the physical 4D KK gravitons viewed on the brane (see Appendix \ref{appendix} for more details on this 4D effective action).

We highlight that even though the junction condition, Eq.(\ref{bcEm}), is the same as that in the RS2 model for the massless mode, the normalization condition, Eq.(\ref{sp}), is different, which will affect the power spectrum of the gravitational waves on the brane as we will show in the next sections.

\section{Kaluza-Klein modes}\label{KK}

To analyze the 3D KK modes on a de Sitter brane embedded in an AdS$_5$ bulk, it is more convenient to rewrite Eq.(\ref{Em}) in terms of the conformal bulk coordinate $z$. Here, we follow closely the methodology used in Refs.~\cite{Dufaux:2004qs,BouhmadiLopez:2004ax}. We start by defining a new field: $\psi_m(z)=\tilde{n}^{3/2}\mathcal{E}_m(z)$. Then, in terms of this new field $\psi_m(z)$, the field equation (\ref{Em}) can be rewritten as
\begin{equation}
-\mathcal{D}_{(+)}\left[q(z)\mathcal{D}_{(-)}\psi_m(z)\right]=m^2w(z)\psi_m(z),\label{perteqz}
\end{equation}
where the derivative operators $\mathcal{D}_{(\pm)}$ are defined as
\begin{equation}
\mathcal{D}_{(\pm)}\equiv\frac d{dz}\pm\frac32\,\frac1{\tilde{n}}\,\frac{d\tilde{n}}{dz},\label{Dop}
\end{equation}
and the functions $q(z)$ and $w(z)$ are given by
\begin{align}
q(z)=&\left(1-\beta\right)\left\{\theta\left[\epsilon(z_b-z)\right]-\epsilon\theta(-z-z_b)\right\},\\
w(z)=&q(z)+2\left[\gamma r_c-\epsilon\,\beta\mu^{-1}\sqrt{1+x^2}\right]\delta\left(z+\epsilon z_b\right),
\end{align}
where $\theta$ is the Heaviside step function. The boundary (junction) condition has been incorporated in Eq.(\ref{perteqz}) through the functions $q(z)$ and $w(z)$. Notice that the \textit{right} side of the bulk, i.e., with $y\geq0$ in the Gaussian coordinate, corresponds to $-z_b\leq z<0$ in the self-accelerating branch ($\epsilon=+1$) and $z_b\leq z$ in the normal branch ($\epsilon=-1$) [cf. Eq.(\ref{z})], respectively.

For $|z|\neq z_b$, after substituting the warp factor (\ref{nz}) in the field equation (\ref{perteqz}), we obtain a Schr\"{o}dinger-like wave equation for $\psi_m$,
\begin{equation}
-\mathcal{D}_{(+)}\mathcal{D}_{(-)}\psi_m=\left[-\frac{d^2}{dz^2}+\bar{V}(z)\right]\psi_m=m^2\psi_m,\label{waveeq}
\end{equation}
where the potential $\bar{V}(z)$ is
\begin{equation}
\bar{V}(z)=\frac{15}4\frac{H^2}{\sinh^2Hz}+\frac94H^2,\label{V}
\end{equation}
and the boundary condition for $\psi_m$ at the brane, i.e., $y\rightarrow0^+$, in the conformal bulk coordinate reads
\begin{equation}
\mathcal{D}_{(-)}\psi_m(-\epsilon z_b)=-\frac{m^2}{1-\beta}\left[\gamma r_c-\epsilon\,\beta\mu^{-1}\sqrt{1+x^2}\right]\psi_m(-\epsilon z_b).\label{bcpsi}
\end{equation}
In addition, the scalar product of eigenmodes defined in Eq.(\ref{sp}) could be expressed in terms of $\psi_m$ and the conformal bulk coordinate as
\begin{align}
\left(\psi_{m},\psi_{\tilde{m}}\right)=&\int_{\mathrm{bulk}}dz w \,\psi_m\psi_{\tilde{m}}\notag\\
=&2\left(1-\beta\right)\int^{v_{\epsilon}}_{-\epsilon z_b} dz \psi_{m}\psi_{\tilde{m}}\notag\\
&+2\left[\gamma r_c-\epsilon\,\beta\mu^{-1}\sqrt{1+x^2}\right]\psi_m(-\epsilon z_b)\psi_{\tilde{m}}(-\epsilon z_b)\notag\\
=&\delta\left(m,\tilde{m}\right),\label{spz}
\end{align}
where $v_+=0$ and $v_-=\infty$.

We now start to investigate the zero-mode (massless mode) $\psi_0$ which, as mentioned in Sec.~\ref{bulkpert}, has the same boundary condition as the one in the RS2 case but not the same normalization. From Eq.(\ref{Em}), a constant solution is found for $\mathcal{E}_0$ in both branches: $\mathcal{E}_0=C_{\epsilon}$, due to the boundary condition imposed at the brane. Therefore, the zero-mode $\psi_0$ reads
\begin{equation}
\psi_0(z)=C_{\epsilon}\,\tilde{n}^{3/2}(z).\label{zeromode}
\end{equation}
The constant $C_{\epsilon}$ will be determined by the normalization condition in Eq.(\ref{spz}): $\left(\psi_0,\psi_0\right)=1$. On the self-accelerating branch, the zero mode is not normalizable since $\psi_0\rightarrow\infty$ as $z\rightarrow0$, leading to a divergent integration in Eq.(\ref{spz}). Thus, the zero mode is not physically allowed in the self-accelerating branch. Conversely, a normalizable zero-mode exists on the normal branch, where we further impose the RS fine-tuning condition \cite{Dufaux:2004qs,BouhmadiLopez:2012uf}
\begin{equation}
\kappa_5^2\lambda=2\mu(3-\beta),\label{finetuning}
\end{equation}
such that the effective cosmological constant on the brane is zero, and most importantly we recover the RS2 model when switching off the GB and IG effects. This will allow us later to compare our results with those of the RS2 model. Moreover, from the modified Friedmann equation at the low energy limit (as $\rho\rightarrow0$), we recover the standard 4D relativistic behavior, with an effective 4D gravitational constant related to the 5D one as follows \cite{BouhmadiLopez:2012uf}:
\begin{equation}
\kappa_4^2=\left(\frac{1-\gamma}{1+\beta}\right)\mu\kappa_5^2,\label{Geff}
\end{equation}
which implies that the IG strength $\gamma$ is bounded: $0\leq\gamma<1$. Making use of the relation (\ref{Geff}), we can reexpress the boundary condition (\ref{bcpsi}) in the normal branch ($\epsilon=-1$) as follows:
\begin{align}
\mathcal{D}_{(-)}\psi_m(z_b)=&-\frac{m^2}{\mu}\bigg[\frac{\gamma}{2(1-\gamma)}\left(\frac{1+\beta}{1-\beta}\right)+\notag\\
&\left(\frac{\beta}{1-\beta}\right)\sqrt{1+x^2}\bigg]\psi_m(z_b),\label{bcpsi2}
\end{align}
which is consistent with the results in the RS2 model modified by only a GB correction ($\gamma\rightarrow0$) \cite{Dufaux:2004qs} and an IG effect alone ($\alpha\rightarrow0$) \cite{BouhmadiLopez:2004ax}. After taking into account the conditions (\ref{finetuning}) and (\ref{Geff}), the scalar product (\ref{spz}) fixs the normalization constant $C_-$ for the massless mode as
\begin{align}
C_-^{-2}=&\frac{(1+\beta)}{\mu}\left[\sqrt{1+x^2}-\left(\frac{1-\beta}{1+\beta}\right)x^2\mathrm{arcsinh}\frac1x\right.\notag\\
         &+\left.\left(\frac{\gamma}{1-\gamma}\right)\right],\label{C-}
\end{align}
which recovers the RS2 result as $\alpha\rightarrow0$ and $\gamma\rightarrow0$ \cite{Langlois:2000ns}, and reduces to the RS2 model modified by either just a GB bulk correction (when $\gamma\rightarrow0$) \cite{Dufaux:2004qs} or an IG brane effect alone (when $\alpha\rightarrow0$) \cite{BouhmadiLopez:2004ax}. From now on, we will focus on KK gravitons within the normal branch.

For a de Sitter brane in the normal branch, the potential (\ref{V}) is a constant at infinity ($|z|\rightarrow\infty$), i.e., $\bar{V}(z\rightarrow\pm\infty)\simeq9/4\,H^2$, which results in a threshold between the continuous and the discrete spectrum of the KK modes. For the heavy modes, with mass scale $m^2>9/4\,H^2$, the asymptotic solutions of the wave equation (\ref{waveeq}) is an oscillating function as $|z|\rightarrow\infty$. Therefore, these heavy modes correspond to the continuous spectrum and are normalizable as plane waves. On the other hand, the asymptotic solutions for the light modes, with $m^2<9/4\, H^2$, contain in general exponentially growing and decaying modes at infinity, in which only the asymptotic decreasing modes, picked by the junction condition, are normalizable. As we discussed earlier, the zero-mode $\psi_0$ in the normal branch is a normalizable solution, but we still need to check whether there exist normalizable light modes other than the zero mode.

%%%%%%
\begin{figure*}[!ht]
  \centering
  \subfloat[]{\label{Fgamma}\includegraphics[width=0.46\textwidth]{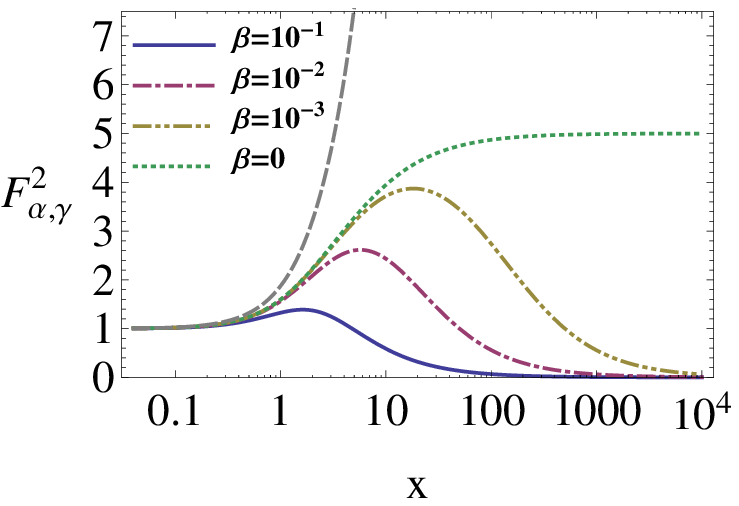}} \!\!\!\!\!
  \subfloat[]{\label{Fbeta}\includegraphics[width=0.46\textwidth]{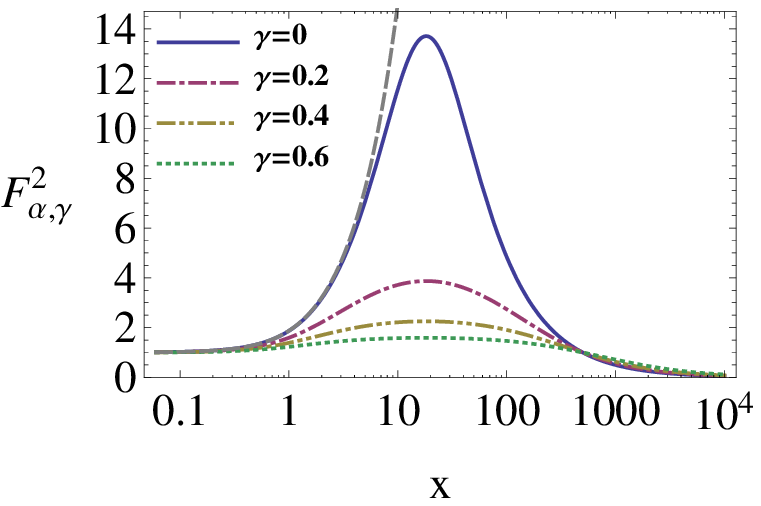}}
  \caption{The correction to the standard 4D amplitude of the tensor perturbations $F^2_{\alpha,\gamma}$ against the dimensionless energy scale of inflation $x\equiv H/\mu$. The dashed-gray line in both plots corresponds to the RS2 model. In (a), we fix the IG strength as $\gamma=0.2$ and change the GB coupling $\beta\equiv4\alpha\mu^2$ as shown on the plot; in (b), we fix the GB coupling as $\beta=10^{-3}$ instead, and vary the IG strength $\gamma$.}
  \label{Fcorrection}
\end{figure*}
%%%%%%%%

To examine the existence of normalizable light modes besides the zero-mode $\psi_0$, it is convenient to introduce new modes defined by \cite{Dufaux:2004qs,BouhmadiLopez:2004ax}
\begin{equation}
\Psi_m(z)=\mathcal{D}_{(-)}\psi_m(z),
\end{equation}
which are the partners of the modes $\psi_m$ in supersymmetric quantum mechanics \cite{Cooper:1994eh}. They have the same spectrum except for the zero-mode $\Psi_0$. Indeed, $\Psi_0=0$, which can be verified using Eqs.(\ref{Dop}) and (\ref{zeromode}). When $|z|\neq z_b$, the wave equation for $\Psi_m$ is obtained by applying the derivative operator $\mathcal{D}_{(-)}$ on Eq.(\ref{waveeq}):
\begin{align}
-\mathcal{D}_{(-)}\mathcal{D}_{(+)}\Psi_m=&\left[-\frac{d^2}{dz^2}+\frac34\frac{H^2}{\sinh^2Hz}+\frac94H^2\right]\Psi_m\notag\\
                                 =&\,m^2\Psi_m.\label{waveeq2}
\end{align}
Supposing that $\beta\neq0$ and $\gamma\neq0$, the asymptotic boundary behavior for $\Psi_m$ near the brane ($z\rightarrow z_b$) is then derived from Eqs.(\ref{waveeq}) and (\ref{bcpsi2}),
\begin{align}
\frac{d\Psi_m}{dz}(z_b^+)=&\bigg\{\frac32\sqrt{H^2+\mu^2}+\mu\left[\frac{\gamma}{2(1-\gamma)}\left(\frac{1+\beta}{1-\beta}\right)
\right.\notag\\
                          &\left.+\left(\frac{\beta}{1-\beta}\right)\sqrt{1+x^2}\right]^{-1}\bigg\}\Psi_m(z_b^+),\label{bcPsi}
\end{align}
where the conditions (\ref{finetuning}) and (\ref{Geff}) have already been taken into account. Specifically, the boundary condition for $\beta=0$ and $\gamma=0$ is given by $\Psi_m(z_b)=0$ [see Eq.(\ref{bcpsi2})]. The advantage is that, unlike the boundary condition for $\psi_m$ (\ref{bcpsi2}), this boundary condition no longer involves the mass dependence of each mode. By operating the integral $\int^{\infty}_{z_b^+}dz\Psi_m\times$ on both sides of Eq.(\ref{waveeq2}), we obtain the following relation:
\begin{align}
\left(m^2-\frac94H^2\right)&\int^{\infty}_{z_b^+}dz\Psi_m^2=\frac34H^2\int^{\infty}_{z_b^+}dz\frac{\Psi^2_m}{\sinh^2Hz}\notag\\
&\!\!\!+\int^{\infty}_{z_b^+}dz\left(\partial_z\Psi_m\right)^2-\left.\Psi_m\partial_z\Psi_m\right|^{\infty}_{z_b^+}.\label{check}
\end{align}
If we expect that the light modes $\psi_m$ with $m^2<9/4\,H^2$ decrease exponentially at infinity, its partners $\Psi_m$ should have the same behavior. Therefore, under this consideration, only the last term evaluated at the lower bound contributes. Moreover, since the GB and IG parameters are bounded as $0\leq\beta<1$ and $0\leq\gamma<1$, the last term at $z=z_b^+$ takes a non-negative value as can be seen from Eq.(\ref{bcPsi}) and vanishes when $\beta=0$ and $\gamma=0$. As a result, the right-hand side of Eq.(\ref{check}) is always positive except for the zero mode, but the left-hand side is negative for the light modes within $0<m^2<9/4\,H^2$ instead. This contradiction leads to the conclusion that the normalizable zero mode with $m^2<9/4\,H^2$ is the only physically admissible light mode.

Consequently, the KK spectrum of the 3D tensor perturbations $h_{ij}$ on the normal branch consists of a massless zero mode and a continuum of heavy modes with $m^2>9/4\,H^2$, which is exactly the same as those in the RS2 model \cite{Langlois:2000ns} and also for the normal branch with an IG term on the brane \cite{BouhmadiLopez:2004ax}. In addition, because of the de Sitter symmetry of the background, the general massive 4D spin-2 gravitons have the same spectrum as that of the 3D tensor modes. Therefore, the spectrum on the normal branch in de Sitter space-time is void of massive ghost states with $0<m^2<2H^2$ \cite{Higuchi:1986py}.

\section{Primordial gravitational waves}\label{priGW}

Here we consider the extreme slow-roll inflation driven by a single inflaton field $\phi$ confined on the brane and study the lowest order primordial fluctuations quantum-mechanically generated during inflation.

The scalar perturbations in this framework have been studied in Ref.~\cite{BouhmadiLopez:2012uf}. The amplitude of the scalar perturbations $A^2_S$ is given by \cite{hep-ph/9912464,Lidsey:1995np,astro-ph/0003278}
\begin{equation}
A^2_S=\frac{9}{25\pi^2}\frac{H^6}{V^2_{\phi}}, \label{AS1}
\end{equation}
where $\phi$ is the inflaton field and $V(\phi)$ is its potential, and we use a subscript $\phi$ to denote a derivative with respect to $\phi$. To compare with the standard 4D general relativity result for a given potential $V$, we can express Eq.(\ref{AS1}) as
\begin{equation}
A^2_S=\frac{\kappa^6_4}{75\pi^2}\left(\frac{V^3}{V_{\phi}^2}\right)G^2_{\alpha,\gamma}(x)=[A^2_S]_{\mathrm{4D}}G^2_{\alpha,\gamma}(x),\label{AS2}
\end{equation}
where $[A^2_S]_{\mathrm{4D}}$ is the standard 4D result. The function $G^2_{\alpha,\gamma}(x)$ is the correction to the standard 4D result, and its explicit form is given in Ref.~\cite{BouhmadiLopez:2012uf}.

In this section, we will analyze the primordial gravitational waves generated in a quasi--de Sitter inflation, where we focus on the 3D tensor modes as discussed in the previous sections.

\subsection{Tensor perturbations on the brane}

Having the KK spectrum analyzed on a de Sitter brane embedded in an AdS$_5$ bulk and described by the action (\ref{action}), we can now calculate the amplitude of the tensorial perturbations generated during an extreme slow-roll inflationary era on the normal branch. From a 4D brane's point of view, we simply treat each KK normalizable mode $\psi_m$ as a quantum field living on a 4D hypersurface (for a 5D viewpoint see, for example, Refs.~\cite{Gorbunov:2001ge,Kobayashi:2003cn,Kobayashi:2005dd}).

When stretched to superhorizon scales during the inflation era, the heavy modes, with $m^2>9/4\,H^2$, continue to oscillate, but their amplitudes are rapidly and strongly suppressed on large scales and remain in their vacuum state, whose contributions can be neglected later on \cite{Langlois:2000ns,Dufaux:2004qs,BouhmadiLopez:2004ax}. By contrast, the zero-mode $\psi_0$ becomes overdamped and approaches a constant on superhorizon scales, which turn into classical perturbations beyond the Hubble radius \cite{Langlois:2000ns,Dufaux:2004qs,BouhmadiLopez:2004ax}.

The second-order effective 4D action for tensor perturbations is given in Eq.(\ref{effaction}). For the massless zero mode, the effective 4D action (\ref{effaction}) with $m=0$ is similar to that for the gravitons in standard 4D general relativity \cite{Lidsey:1995np} up to a renormalization factor: $\bar{h}^{(0)}_{ij}=C_-(\kappa_5/\kappa_4)\,h^{(\mathrm{4D})}_{ij}$, where $h^{(\mathrm{4D})}_{ij}$ is the standard 4D gravitons, and it will therefore rescale the amplitude of the corresponding tensor perturbations in the standard form. With this on hand, the normalized amplitude of the tensor perturbations $A^2_T$, following the notation used in Ref.~\cite{Lidsey:1995np}, is then given by
\begin{equation}
A^2_T=\frac{2\kappa_4^2}{25}\left(\frac H{2\pi}\right)^2F^2_{\alpha,\gamma}=[A^2_T]_{\mathrm{4D}}F^2_{\alpha,\gamma},\label{AT}
\end{equation}
where $[A^2_T]_{\mathrm{4D}}$ is the standard form of 4D GR, and $F^2_{\alpha,\gamma}$ is the correction to this standard form,
\begin{equation}
F^2_{\alpha,\gamma}=C_-^2\,\left(\frac{\kappa_5}{\kappa_4}\right)^2.\label{Fc}
\end{equation}
Substituting the relation (\ref{Geff}) and the normalized constant (\ref{C-}) into the correction factor given in Eq.~(\ref{Fc}), we can rewrite it as a function of $x$ with the GB and the IG couplings,
\begin{equation}
F^{-2}_{\alpha,\gamma}(x)=\gamma+(1-\gamma)\left[\sqrt{1+x^2}-\left(\frac{1-\beta}{1+\beta}\right)x^2 \mathrm{arcsinh}\frac1x\right].\label{F}
\end{equation}
This result reduces to the RS2 case \cite{Langlois:2000ns} when $\alpha\rightarrow0$ and $\gamma\rightarrow0$; moreover, it is consistent with the result in a modified RS2 model by a GB correction \cite{Dufaux:2004qs} or an IG correction \cite{BouhmadiLopez:2004ax}. The behavior of the correction factor $F^2_{\alpha,\gamma}(x)$ is qualitatively the same as the one for the scalar perturbations in the same system \cite{BouhmadiLopez:2012uf}.

%%%%%%
\begin{figure*}[!ht]
  \centering
  \subfloat[]{\label{Rgamma}\includegraphics[width=0.46\textwidth]{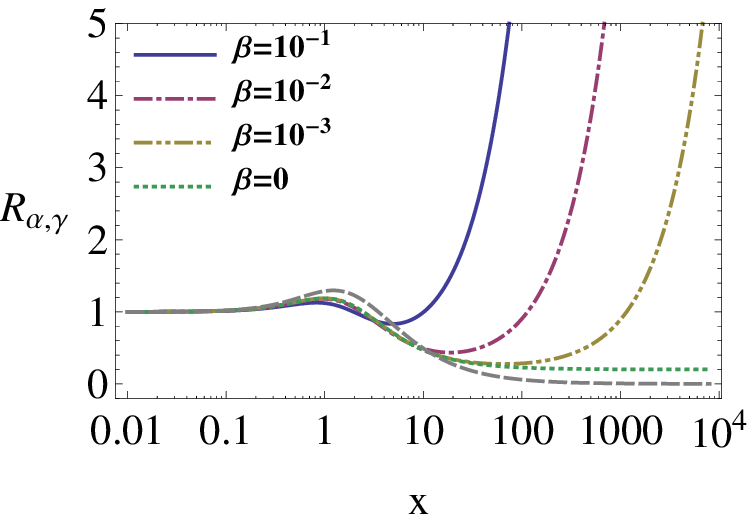}} \!\!\!\!\!
  \subfloat[]{\label{Rbeta}\includegraphics[width=0.46\textwidth]{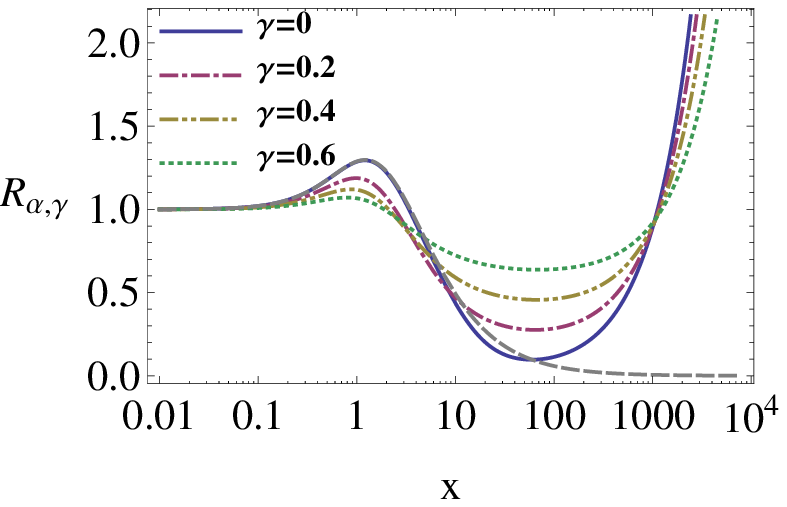}}
  \caption{The correction to the standard 4D tensor-to-scalar ratio $R_{\alpha,\gamma}$ against the dimensionless energy scale of inflation $x\equiv H/\mu$. The dashed-gray line in both figures corresponds to the RS2 model. In (a), we have fixed the IG strength as $\gamma=0.2$ and changed the GB coupling $\beta\equiv4\alpha\mu^2$ as shown on the plot; in (b), we have fixed the GB coupling as $\beta=10^{-3}$ and changed the IG strength $\gamma$.}
  \label{Rcorrection}
\end{figure*}
%%%%%%%%

In the relatively low energy regime as compared with the AdS$_5$ energy scale, i.e., when $H\ll\mu$ or $x\rightarrow0$, the correction fulfills $F^2_{\alpha,\gamma}(x)\approx1$, recovering therefore the standard 4D result; while at relatively high energy as $H\gg\mu$ or $x\gg1$, for $\beta\neq0$ and $\gamma\neq0$ we obtain
\begin{equation}
F^2_{\alpha,\gamma}(x)\approx\frac{1+\beta}{2\beta(1-\gamma)}\,\frac1x.
\end{equation}
Consequently, the amplitude of the tensor perturbations will be suppressed by the GB effect when inflation takes place at relatively high energies, i.e., as $H\gg\mu$. The behaviors of the correction (\ref{F}) with respect to the dimensionless energy scale $x$, for different GB and IG couplings, are shown in Fig.~\ref{Fcorrection}. The dashed-gray line in Fig.~\ref{Fcorrection} corresponds to the RS2 model, which is a monotonically increasing function \cite{Langlois:2000ns}. However, unlike the RS2 model, both GB and IG effects tend to suppress the RS2 enhancement. Furthermore, the larger the GB and IG effects, the stronger the suppression becomes. In the intermediate energy regime, the correction $F^2_{\alpha,\gamma}(x)$ with $\beta\neq0$ will be enhanced up to a maximum value as the dimensionless energy scale $x$ increases, and then will be suppressed by the GB effect at higher energies (see Fig.~\ref{Fcorrection}), independently of the IG strength. The maximum value of the correction $F^2_{\alpha,\gamma}$ is given by
\begin{equation}
(F^2_{\alpha,\gamma})_{\mathrm{m}}=\left[\gamma+(1-\gamma)\frac{1+\beta(1+x^2_{\mathrm{m}})}{(1+\beta)\sqrt{1+x^2_{\mathrm{m}}}}\right]^{-1},\label{Fm}
\end{equation}
where the critical energy scale $x_{\mathrm{m}}$ is determined by the following relation for a given value of $\beta$:
\begin{equation}
\sqrt{1+x^2_{\mathrm{m}}}\,\,\mathrm{arcsinh}\frac1{x_{\mathrm{m}}}=\frac1{1-\beta}.\label{xm}
\end{equation}
The above relation (\ref{xm}) gives a one-to-one correspondence between the critical energy scale $x_{\mathrm{m}}$ and the GB coupling $\beta$. Moreover, it can be shown that the critical energy scale $x_{\mathrm{m}}$ will decrease as the GB coupling $\beta$ increases [cf.~Eq.(\ref{xm}) and Fig.~\ref{Fcorrection}]. In addition, the physically allowed region for a nonvanishing GB coupling, $0<\beta<1$, covers the region $0<x_{\mathrm{m}}$ for the critical energy scale. Therefore, as the GB coupling $\beta$ increases, the maximum correction $(F^2_{\alpha,\gamma})_{\mathrm{m}}$ given in Eq.(\ref{Fm}) decreases, and furthermore, it can be shown that the maximum correction $(F^2_{\alpha,\gamma})_{\mathrm{m}}$ is bounded as $1<(F^2_{\alpha,\gamma})_{\mathrm{m}}<1/[\gamma+\beta(1-\gamma)]$. The limiting maximum value of the previous inequality is chosen to clearly imply that the maximum correction $(F^2_{\alpha,\gamma})_{\mathrm{m}}$ is always finite unless both GB parameter $\beta$ and IG strength $\gamma$ vanish simultaneously, which is the RS2 case. A similar behavior has been found for the amplitude of the scalar perturbations \cite{BouhmadiLopez:2012uf}. The amplitude in the modified RS2 model with only the GB effect has qualitatively similar behavior but without the IG suppression on the maximum correction $(F^2_{\alpha,\gamma})_{\mathrm{m}}$ \cite{Dufaux:2004qs}; while in the RS2 model modified by just the IG effect, the RS2 enhancement at high energies is bounded and approaches the constant value $1/\gamma$ \cite{BouhmadiLopez:2004ax}.

\subsection{The tensor-to-scalar ratio}

%%%%%%
\begin{figure*}[!ht]
  \centering
  \subfloat[]{\label{Qgamma}\includegraphics[width=0.46\textwidth]{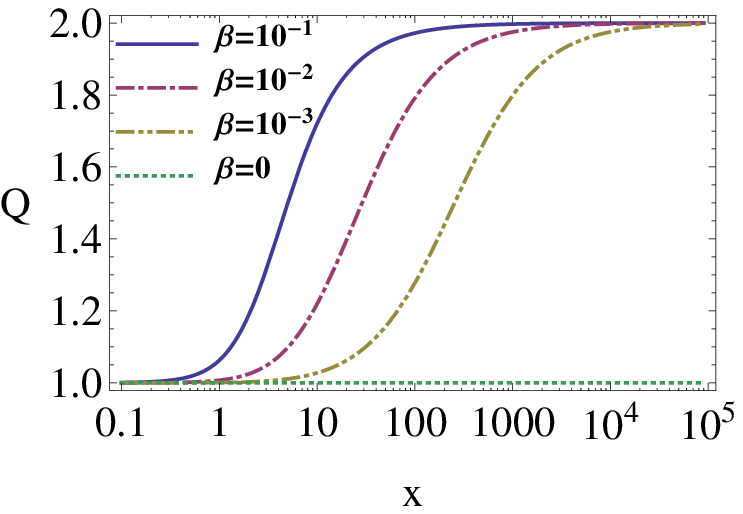}} \!\!\!\!\!
  \subfloat[]{\label{Qbeta}\includegraphics[width=0.46\textwidth]{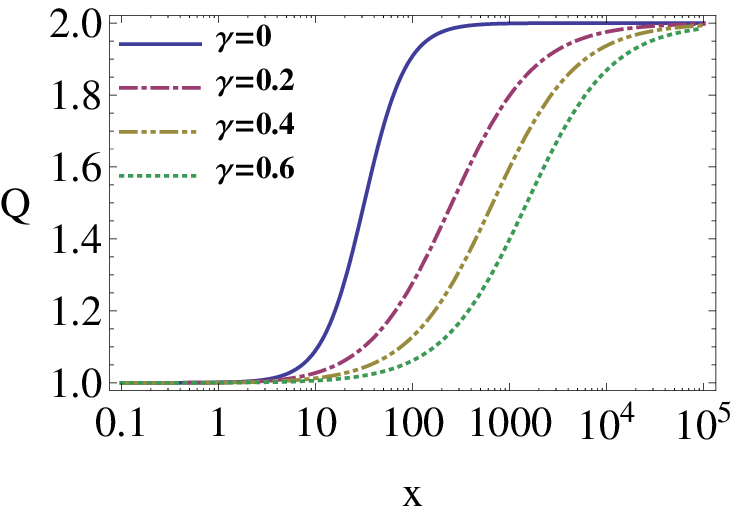}}
  \caption{The parameter $Q$ against the dimensionless energy scale $x$. In (a), we have fixed the IG strength as $\gamma=0.2$ and changed the GB coupling $\beta\equiv4\alpha\mu^2$ as shown on the plot; in (b), we have fixed the GB coupling as $\beta=10^{-3}$ and changed the IG strength $\gamma$. The RS2 case corresponds to a horizontal line with $Q=1$. A deviation of $Q$ from unity implies the nonfulfillment of the standard consistency relation.}
  \label{Qshift}
\end{figure*}
%%%%%%%%

During a quasi--de Sitter brane inflation with the GB and IG modifications, as compared with the standard 4D case, the amplitudes of both scalar and tensor perturbations are enhanced at intermediate energy scales. Conversely, as the inflationary energy scale gets very high, the amplitudes will be highly suppressed by the GB effect (see Fig.~\ref{Fcorrection} and Ref.~\cite{BouhmadiLopez:2012uf}). At such high energies as $x\gg1$, the correction to the standard amplitude of the scalar perturbations $G^2_{\alpha,\gamma}(x)$ (with $\beta\neq0$ and $\gamma\neq0$) is given by \cite{BouhmadiLopez:2012uf}
\begin{equation}
G^2_{\alpha,\gamma}(x)\approx\frac{27}{64}\left[\frac{1+\beta}{2\beta(1-\gamma)}\right]^3\,\frac1{x^3}.\label{Gh}
\end{equation}
The scalar suppression shown in Eq.(\ref{Gh}) is much stronger than the one corresponding to the tensor sector [cf.~(\ref{F})]. Therefore, the tensor-to-scalar ratio $A^2_T/A^2_S$ will be enhanced at very high energies. The general form of the tensor-to-scalar ratio can be written as
\begin{equation}
\frac{A^2_T}{A^2_S}=\epsilon_{\mathrm{4D}}\,R_{\alpha,\gamma}(x),
\end{equation}
where $\epsilon_{\mathrm{4D}}$ is the standard 4D slow-roll parameter $\epsilon_{\mathrm{4D}}\equiv1/2\kappa_4^2\times(V_{\phi}/V)^2$, which corresponds to the standard GR result for the ratio between the amplitude of the tensor and the scalar modes, and the correction to the standard 4D tensor-to-scalar ratio $R_{\alpha,\gamma}$ is given by
\begin{equation}
R_{\alpha,\gamma}(x)=\frac{3\,x^2}{2(1-\gamma)}\left(\frac{1+\beta}{3-\beta}\right)\left(\frac{\lambda}{V}\right)\frac{F^2_{\alpha,\gamma}(x)}
{G^2_{\alpha,\gamma}(x)}\,.
\end{equation}
The ratio $V/\lambda$ can be further expressed as a function of $x$ as well as the GB and the IG parameters $\beta$ and $\gamma$ by using Eqs.(\ref{Friedmann}), (\ref{finetuning}), and (\ref{Geff}), and the explicit form of $G^2_{\alpha,\gamma}(x)$ was obtained in Ref.~\cite{BouhmadiLopez:2012uf}. The qualitative behavior of $R_{\alpha,\gamma}$ is shown in Fig.~\ref{Rcorrection}. The GB effect, as can be seen from Fig.~\ref{Rcorrection}, will significantly affect the correction $R_{\alpha,\gamma}$ in the relatively high-energy regime, which is already the case in the absence of any IG effect \cite{Dufaux:2004qs}. The IG effect, on the other hand, has a tendency to counterbalance the abrupt enhancement of $R_{\alpha,\gamma}$ caused by the GB effect at high energies. Furthermore, the larger is the IG strength (larger $\gamma$), the stronger will be the counterbalanced effect. However, the IG effect cannot avoid the unbounded enhancement caused by the GB effect at very high energies, at least within the ``slim'' brane approximation we have used. Consequently, observational constraints on the production of the gravitons in the early universe will strongly limit the energy scale where inflation takes place.

For a standard single-field slow-roll inflation, the tensor-to-scalar ratio in the slow-roll approximation can be expressed in terms of the consistency relation,
\begin{equation}
\frac{A^2_T}{A^2_S}=-\frac12n_T,\label{consistency}
\end{equation}
where the tensor spectral index $n_T$ is defined by $n_T\equiv d\ln A^2_T/d\ln k$. This consistency relation relates observable quantities in the inflationary paradigm, and therefore it provides an observational test for the standard single-field inflation. In addition, the consistency relation (\ref{consistency}) holds as well for other modified gravity theories \cite{Tsujikawa:2004my,Huey:2001ae,Huey:2002xw,Giudice:2002vh,Seery:2003ge}, and it is important to examine whether or not the GB and IG effects would break the consistency relation (\ref{consistency}).

We will now proceed to investigate the consistency relation in this system. In the extreme slow-roll approximation, the tensor spectral index $n_T$ can be written as
\begin{equation}
n_T\equiv\left.\frac{d\ln A^2_T}{d\ln k}\right|_{k=aH}\simeq-\frac{d\ln\left(HF_{\alpha,\gamma}\right)^{-2}}{d\ln H}\cdot\frac{d\ln H}{d\ln a},\label{nT}
\end{equation}
where we have substituted the relation (\ref{AT}) into the definition of $n_T$. To conveniently compare the tensor-to-scalar ratio with the expression (\ref{nT}), we replace $V_{\phi}$ in Eq.(\ref{AS1}) using the field equation of the inflaton field in the slow-roll limit: $3H\dot{\phi}\simeq-V_{\phi}$, which can be rewritten as
\begin{equation}
V_{\phi}^2=-3H^3\frac{dV}{dH}\frac{d\ln H}{d\ln a}.
\end{equation}
Substituting this relation in the amplitude of scalar perturbations (\ref{AS1}), we have
\begin{equation}
A^2_S=-\frac{3}{25\pi^2}H^3\frac{dH}{dV}\left(\frac{d\ln H}{d\ln a}\right)^{-1}.\label{AS3}
\end{equation}
Then, by putting together Eqs.(\ref{AT}), (\ref{nT}), and (\ref{AS3}), we obtain a simplified form for the tensor-to-scalar ratio,
\begin{equation}
\frac{A^2_T}{A^2_S}=-\frac Q2n_T,\label{ratio}
\end{equation}
where the parameter $Q$ is given by
\begin{equation}
Q=-\frac{\kappa_4^2}{3H}\frac{dV}{dH}F^2_{\alpha,\gamma}\left[\frac{d\ln \left(HF_{\alpha,\gamma}\right)^{-2}}{d\ln H}\right]^{-1}.\label{Q}
\end{equation}
From Eqs.(\ref{ratio}) and (\ref{Q}), we can see that only when $Q=1$ does the consistency relation hold in this system. The parameter $Q$ is therefore a crucial factor to be analyzed.

The expression for $Q$ in Eq.(\ref{Q}) can be further simplified as follows. Indeed, an important property for the correction $F^2_{\alpha,\gamma}$ in Eq.(\ref{F}) is that it satisfies a first-order differential equation,
\begin{align}
\frac{d}{d\ln x}\left[\ln\left(xF_{\alpha,\gamma}\right)^{-2}\right]&=-2F_{\alpha,\gamma}^2\bigg[\gamma+(1-\gamma)\notag\\
&\times\frac{1+\beta(1+x^2)}{(1+\beta)\sqrt{1+x^2}}\bigg],\label{r1}
\end{align}
which is consistent with the results in the generalized RS2 model with just a GB correction ($\gamma\rightarrow0$) \cite{Dufaux:2004qs} and an IG effect alone ($\alpha\rightarrow0$) \cite{BouhmadiLopez:2004ax}. Furthermore, using the modified Friedmann equation (\ref{Friedmann}) (for the normal branch $\epsilon=-1$) as well as the conditions (\ref{finetuning}) and (\ref{Geff}), we obtain the following relation:
\begin{equation}
\left(H\frac{dH}{dV}\right)^{-1}=\frac6{\kappa_4^2}\left[\gamma+(1-\gamma)\,\frac{1+\beta(1+2x^2)}{(1+\beta)\sqrt{1+x^2}}\right].\label{r2}
\end{equation}
The relations (\ref{r1}) and (\ref{r2}) can then be employed to simplify the parameter $Q$ in Eq.(\ref{Q}). As a result, the parameter $Q$ becomes a simpler function of the dimensionless energy scale $x$ as well as the GB and the IG parameters $\beta$ and $\gamma$,
\begin{equation}
Q=\frac{(1-\gamma)+\beta(1-\gamma)(1+2x^2)+\gamma(1+\beta)\sqrt{1+x^2}}{(1-\gamma)+\beta(1-\gamma)(1+x^2)+\gamma(1+\beta)\sqrt{1+x^2}}.\label{Q2}
\end{equation}
The previous function reduces to the results of the RS2 model with just a GB effect ($\gamma\rightarrow0$) \cite{Dufaux:2004qs} and an IG effect alone ($\alpha\rightarrow0$) \cite{BouhmadiLopez:2004ax}. In principle, the parameter $Q$ (\ref{Q2}) depends on the energy scale of inflation, implying that the GB correction does break the consistency relation. Nevertheless, the relation is not altered abruptly by GB effect: in the relatively low-energy regime (as $x\rightarrow0$), the parameter $Q\rightarrow1$ recovering the standard consistency relation (\ref{consistency}), while the parameter $Q\rightarrow2$ at relatively high energies ($x\gg1$). It can be shown that $Q$ is bounded for any value of $x$ ($1\leq Q<2$), the behavior of which is shown in Fig.~\ref{Qshift}. As we can see in Fig.~\ref{Qshift}, the IG strength tends to lower the value of the parameter $Q$ [see also Eq.(\ref{Q2})], but the GB effect still leads to a deviation from the standard consistency relation (\ref{consistency}) at very high energies, even for a tiny GB correction (cf.~pure GB brane wrold \cite{Dufaux:2004qs}).

\subsection{Observational constraints}

%%%%%%%%%%%%%%%%%%%%%%%%%%%%%%%%%
\begin{figure*}[!ht]
  \centering
  \subfloat[]{\label{Vgamma}\includegraphics[width=0.35\textwidth]{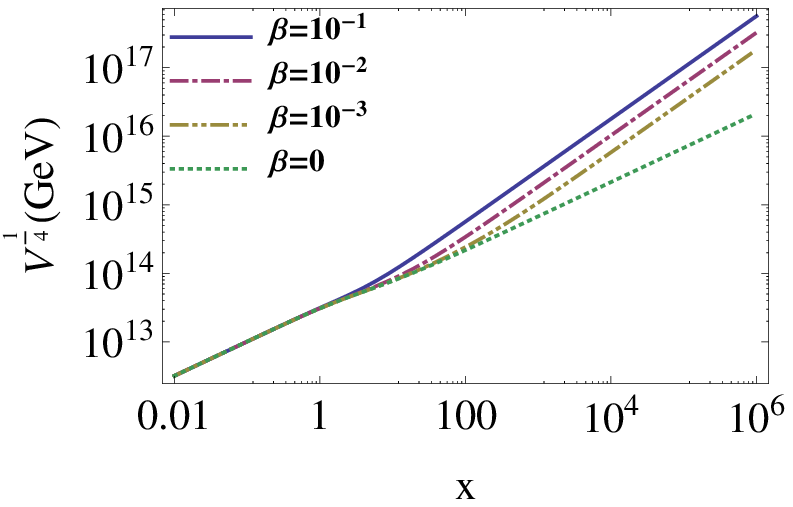}}
  \subfloat[]{\label{epsilongamma}\includegraphics[width=0.35\textwidth]{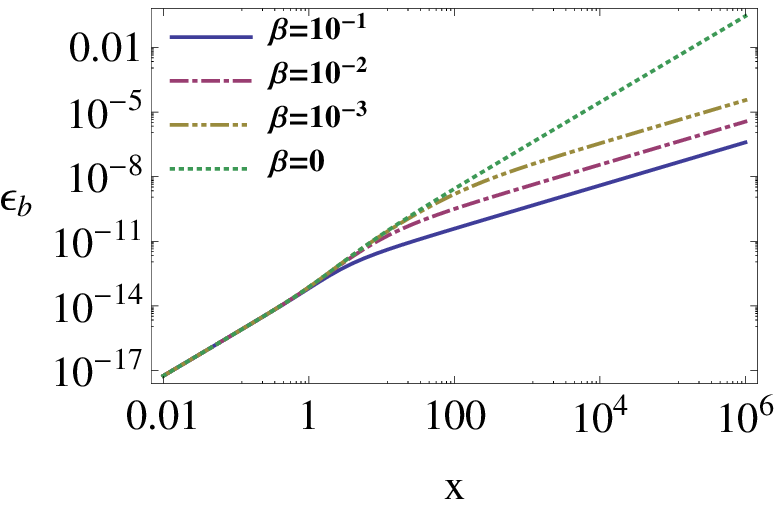}}
  \subfloat[]{\label{ratiogamma}\includegraphics[width=0.35\textwidth]{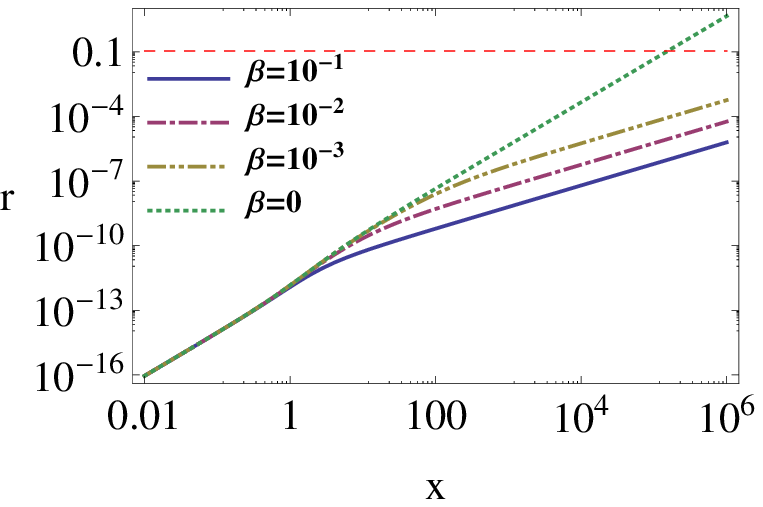}}
  \caption{The energy scale of the inflaton potential (a), the slow-roll parameter $\epsilon_b$ (b), and the tensor-to-scalar ratio $r\equiv P_T/P_S$ (c) versus the dimensionless energy scale $x\equiv H/\mu$. Here we have imposed the constraint on the power spectrum of the scalar perturbations from Planck data: $P_S=2.215\times10^{-9}$. Then, we fix the AdS$_5$ energy scale $\mu$ as $\kappa_4\mu=10^{-10}$ (just as an example) and the IG strength, $\gamma=0.2$, and change the GB parameter $\beta$ as shown on the plots. The horizontal dashed red line in (c) indicates the upper bound for the tensor-to-scalar ratio $r$ predicted by Planck.}
	\label{consistency1}
\end{figure*}
%%%%%%%%%%%%%%%%%%%%%%%%%%%%%%%%%
%%%%%%%%%%%%%%%%%%%%%%%%%%%%%%%%%
\begin{figure*}[!ht]
  \centering
  \subfloat[]{\label{Vgamma}\includegraphics[width=0.35\textwidth]{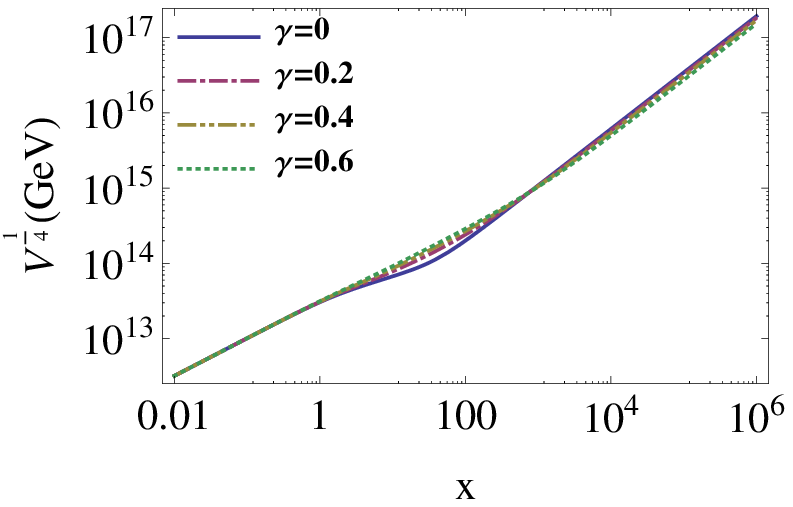}}
  \subfloat[]{\label{epsilongamma}\includegraphics[width=0.35\textwidth]{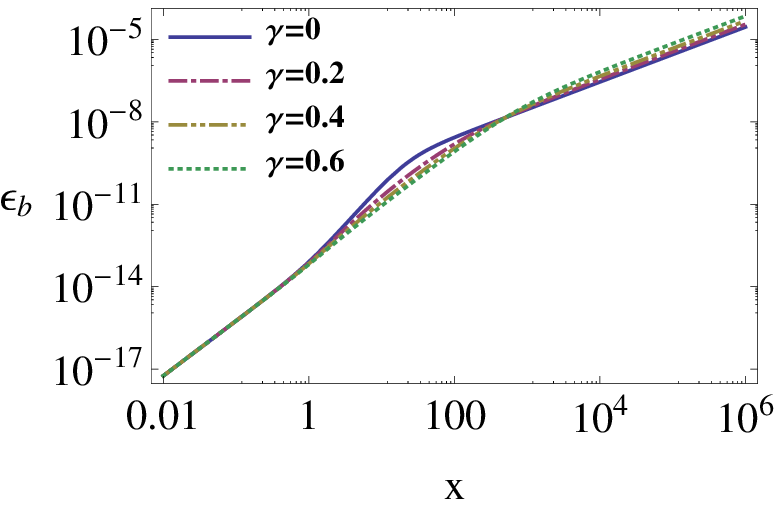}}
  \subfloat[]{\label{ratiogamma}\includegraphics[width=0.35\textwidth]{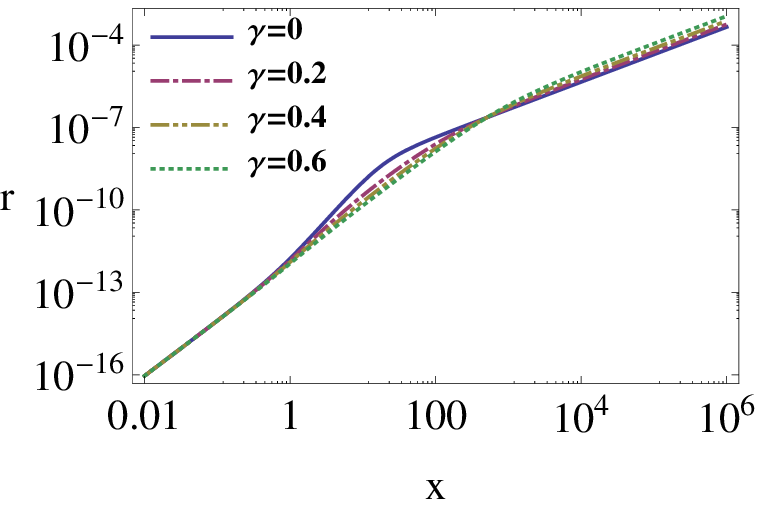}}
  \caption{The energy scale of the inflaton potential (a), the slow-roll parameter $\epsilon_b$ (b), and the tensor-to-scalar ratio $r\equiv P_T/P_S$ (c) versus the dimensionless energy scale $x\equiv H/\mu$. Here we have imposed the constraint on the power spectrum of the scalar perturbations from Planck data: $P_S=2.215\times10^{-9}$. Then, we fix the AdS$_5$ energy scale $\mu$ as $\kappa_4\mu=10^{-10}$ (as an example) and the GB coupling, $\beta=10^{-3}$, and change the IG strength $\gamma$ as shown on the plots.}
	\label{consistency2}
\end{figure*}
%%%%%%%%%%%%%%%%%%%%%%%%%%%%%%%%%

We now discuss the implications of the observational constraints from the Planck mission, which gives a constraint on the power spectrum of the scalar perturbations $P_S$ \cite{Ade:2013zuv} (notice that the amplitudes $A^2_S$ and $A^2_T$ are normalized as $A^2_S\equiv4/25\, P_S$ and $A^2_T\equiv1/100\, P_T$, respectively): $P_S=2.215\times10^{-9}$ at the pivot scale $k_*=0.05\mathrm{Mpc}^{-1}$, and an upper bound on the tensor-to-scalar ratio $r$ \cite{Ade:2013zuv,Ade:2013uln}: $r\equiv P_T/P_S\leq0.11$ at the pivot scale $k_*=0.002 \,\mathrm{Mpc}^{-1}$. We expect that for these large pivot scales the extreme slow-roll approximation is indeed valid, because these are the modes that exit the horizon well inside the inflationary era.

The spectral index $n_S$ for the scalar perturbations, within the extreme slow-roll limit, can be written in terms of the first two slow-roll parameters,
\begin{equation}
n_S\equiv\left.\frac{d\ln A^2_S}{a\ln k}\right|_{k=aH}\simeq1-6\epsilon_b+2\eta_b,
\end{equation}
where the slow-roll parameters $\epsilon_b$ and $\eta_b$ are defined as
\begin{align}
\epsilon_b\equiv & -\frac{\dot{H}}{H^2}=\frac1{2\kappa_4^2}\left(\frac{V_{\phi}}{V}\right)^2 C^{\,(1)}_{\alpha,\gamma}(x),\\
\eta_b\equiv & \,\,\frac{V_{\phi\phi}}{3H^2}=\frac1{\kappa_4^2}\left(\frac{V_{\phi\phi}}{V}\right) C^{\,(2)}_{\alpha,\gamma}(x),
\end{align}
where $C^{\,(1)}_{\alpha,\gamma}$ and $C^{\,(2)}_{\alpha,\gamma}$ are the corrections to the standard 4D expressions, the exact forms of which have been obtained for the brane-world model in Ref.~\cite{BouhmadiLopez:2012uf}; moreover, both parameters, during the extreme slow-roll inflation, are expected to be tiny.

To check the consistency between the observations and the predictions of this model, we rewrite the amplitudes $A^2_S$ (\ref{AS2}) and $A^2_T$ (\ref{AT}) in the convenient forms
\begin{align}
A^2_S=&\frac{(\kappa_4\mu)^2}{75\pi^2\epsilon_b}(1-\gamma)\left(\frac{3-\beta}{1+\beta}\right)\left(\frac V{\lambda}\right)C^{\,(1)}_{\alpha,\gamma}\,\,G^2_{\alpha,\gamma},   \label{AS4} \\
A^2_T=&\frac{(\kappa_4\mu)^2}{50\pi^2}x^2 F^2_{\alpha,\gamma}.\label{AT2}
\end{align}
Given that the value of the power spectrum $P_S$ is constrained by Planck data, the slow-roll parameter $\epsilon_b$ and the ratio $r\equiv P_T/P_S$ [cf.~Eqs.(\ref{AS4}) and (\ref{AT2})] can be evaluated as a function of the dimensionless energy scale $x\equiv H/\mu$ for a given set of the parameters $\kappa_4\mu$, $\beta$, and $\gamma$. It is important to check the regime of validity of the slow-roll approximation for the set of parameters mentioned above, as our analysis is based on the correctness of the extreme slow-roll approximation. We have checked our previous work \cite{BouhmadiLopez:2012uf} and found that indeed the slow-roll approximation is valid up to quite a large value of $x$ for a physically reasonable set of the dimensionless parameters $\kappa_4\mu$, $\beta$, and $\gamma$ given the measured amplitude of the scalar perturbations by WMAP7. This conclusion is unaltered once we update our analysis to Planck data for $P_S$ as Figs.~\ref{consistency1} and \ref{consistency2} show. In these figures, we fix AdS$_5$ energy scale $\kappa_4\mu=10^{-10}$ as an example. Within the validity of the extreme slow-roll regime, we can predict the value of the tensor-to-scalar ratio $r$. Our results can be seen in Figs.~\ref{consistency1} and \ref{consistency2}, which show that the amplitude of the tensor perturbations are much smaller than those expected from the scalar perturbations. Therefore, the predicted values of $r$ fulfill by far the bound on this magnitude as imposed by Planck data, except at very large $x$ or equivalently when the energy scale of the brane is much larger than the one of the bulk, and where the extreme slow-roll approximation starts to cease to be valid. It can be shown as well that the energy scale of the potential $V$, the slow-roll parameter $\epsilon_b$, and the tensor-to-scalar ratio $r$ are more sensitive to the GB effect than the IG effect (see Figs.~\ref{consistency1} and \ref{consistency2}).

\section{conclusions}\label{conclusion}

High energy phenomena, like those corresponding to the inflationary era, are the ideal arena to look for the fingerprints of string/M theory. A phenomenological approach to tackle this issue is rooted on the idea of the brane-world where our Universe is described by a brane embedded in an extradimensional space-time, i.e., a bulk. Indeed, brane-world cosmology provides an excellent way to explore the nature of extra dimensions without going into the whole complexity of string/M theory.

In this paper, we have investigated the RS2 kind of the brane-world model; i.e., the bulk corresponds to an AdS$_5$ and the 4D effective cosmological constant is fine-tuned to be zero, with two curvature corrections: a GB contribution to the bulk action and an IG effect to the brane action. Both effects are expected to be enhanced at high energies as is the case during the inflationary era. Within this setup, we have analyzed the primordial gravitational waves, the 3D tensor perturbations, produced during an inflationary era which is driven by an inflaton confined on the brane. Inflation is modeled within the extreme slow-roll approximation; i.e., the brane is described by a de Sitter universe. Although the perturbed bulk field equation of the 3D tensor modes is not modified in the linear order by taking into account the GB and IG effects, these curvature terms modify the boundary conditions at the brane, which is important at relatively high energies where inflation takes place as it affects the amplitude of the tensor perturbations as we have shown.

The mass spectrum of KK modes on the normal branch is composed of a massless zero mode and a continuum of massive KK modes with $m^2>9/4H^2$, which is the same spectrum as that in the RS2 model \cite{Langlois:2000ns}. Since the massive modes are too heavy to be excited during inflation, we can then safely neglect their contributions and focus only on the zero-mode graviton. As a result, the combined effect from both GB and IG curvature modifications, in the relatively high-energy inflationary era, gives rise to a suppression effect on the amplitude of the primordial gravitational waves as compared with the one in the original RS2 model (cf.~the pure RS2 model \cite{Langlois:2000ns}, the modified RS2 model with only a GB effect \cite{Dufaux:2004qs} and an IG effect alone \cite{BouhmadiLopez:2004ax}). Furthermore, the tensorial amplitudes are strongly suppressed by the GB effect at very high energies. These qualitative behaviors are similar to those of the amplitude of the scalar perturbations for the same system \cite{BouhmadiLopez:2012uf}. Additionally, relative to the standard 4D results, the GB effect in particular abruptly increases the tensor-to-scalar ratio and breaks the standard consistency relation in the relatively high-energy regime, which is already the case in the absence of any IG effect \cite{Dufaux:2004qs}. These changes cannot be entirely ameliorated by invoking the IG effect; nevertheless, the IG effect will mildly counterbalance these significant changes caused by the GB effect at high energies.

Finally, we have constrained the model by considering the amplitude of the scalar perturbations $P_S$ and the tensor-to-scalar ratio $r$, recently obtained by the Planck mission \cite{Ade:2013zuv,Ade:2013uln}. More precisely, we have used the power spectrum to pick up the region of validity of the slow-roll approximation, a cornerstone of our approach. Then, for that region we have obtained the predicted tensor-to-scalar ratio $r$ as Figs.~\ref{consistency1} and \ref{consistency2} show for some values of the parameter space. The model is in agreement with Planck data for a very wide range of the energy scale of inflation. However, since $r$ is an increasing function of the energy scale ($x\equiv H/\mu$), this model ceases to be consistent with the observations at very high energies, resulting in an upper bound of the allowed energy scale of inflation on this kind of model for a given region of the parameter space. The region where the extreme slow-roll approximation ceases to be valid, and therefore our analysis, is precisely the region where the predicted $r$ is too large as compared with Planck data.

\acknowledgments
M.B.L. is supported by the Basque Foundation for Science IKERBASQUE. She also wishes to acknowledge the hospitality of LeCosPA Center at the National Taiwan University during the completion of part of this work and the support of the Portuguese Agency ``Funda\c{c}\~{a}o para a Ci\^{e}ncia e Tecnologia" through PTDC/FIS/111032/2009. This work was partially supported by the Basque government Grant No. IT592-13. K.I. is supported by Taiwan National Science Council (TNSC) under Project No. NSC101-2811-M-002-103. P.C. and Y.W.L. are supported by Taiwan National Science Council (TNSC) under Project No. NSC 97-2112-M-002-026-MY3 and by Taiwan National Center for Theoretical Sciences (NCTS). P.C. is in addition supported by U.S. Department of Energy under Contract No. DE-AC03-76SF00515.

%This work has been supported by a Spanish-Taiwanese Interchange Program with reference 2011TW0010 (Spain) and NSC 101-2923-M-002-006-MY3 (Taiwan)

\appendix
\section{THE 4D EFFECTIVE ACTION}\label{appendix}

The 4D effective action for the 3D tensor modes can be determined by introducing an auxiliary first-order matter content on the brane \cite{Izumi:2006ca}. With this matter content restricted to the brane, the full perturbed field equation up to first order reads
\begin{align}
&\left[\partial^2_t+3H\partial_t-e^{-2Ht}\nabla^2-\left(n^2\partial^2_y+4nn'\partial_y\right)\right]h_{ij}\notag\\
&=-\frac2{1-\beta}\bigg[\left(\gamma r_c-\epsilon\,\beta\mu^{-1}\sqrt{1+x^2}\right)\times\notag\\
&\quad\left(\partial^2_t+3H\partial_t-e^{-2Ht}\nabla^2\right)h_{ij}-\kappa^2_5e^{-2Ht}T_{ij}\bigg]\,\delta\left(y\right)\,,\label{perteq2}
\end{align}
where $T_{ij}$ is the energy momentum tensor for the matter content on the brane, which is assumed to be a transverse and traceless source.

After the KK decomposition (\ref{decomposition}), it can be shown, from Eq.(\ref{Em}), that the eigenmodes $\mathcal{E}_{m}$ are mutually orthogonal with respect to the scalar product defined by
\begin{align}
&\int^{u_{\epsilon}}_{-u_{\epsilon}}dyn^2\mathcal{E}_{m}
\mathcal{E}_{\tilde{m}}+\notag\\
&\frac2{1-\beta}\left[\gamma r_c-\epsilon\,\beta\mu^{-1}\sqrt{1+x^2}\right]\mathcal{E}_m(0)\mathcal{E}_{\tilde{m}}(0)\notag\\
&\equiv\chi_m\left(\mathcal{E}_{m},\mathcal{E}_{\tilde{m}}\right)=\chi_m\,\delta\left(m,\tilde{m}\right)\label{sp2},
\end{align}
where the delta function $\delta\left(m,\tilde{m}\right)$ is a Kronecker delta for the discrete modes and a Dirac delta function for the continuous spectrum; $\chi_m$ is an arbitrary constant, indicating that there exists a degree of freedom for the normalization of each mode. Then, integrating out the extra dimension by performing ($\int^{u_{\epsilon}}_{-u_{\epsilon}}dyn^2\mathcal{E}_{m}\times$) on both side of Eq.(\ref{perteq2}) with the decomposition (\ref{decomposition}) and the above scalar product (\ref{sp2}), we obtain the perturbed 4D effective field equation on the brane,
\begin{align}
\chi_m&(\partial^2_t+3H\partial_t-e^{-2Ht}\nabla^2+m^2)\bar{h}^{(m)}_{ij}\notag\\
&-\frac{2\kappa^2_5}{1-\beta}\,e^{-2Ht}\,\mathcal{E}_m^2(0)T_{ij}=0,
\label{4Deffequation}
\end{align}
where $\bar{h}^{(m)}_{ij}=h^{(m)}_{ij}\mathcal{E}_m(0)$, the physical 4D KK gravitons on the brane.

From the 4D effective field equation (\ref{4Deffequation}), the second-order effective action for the KK gravitons $\bar{h}^{(m)}_{ij}$ can be written as
\begin{align}
S_h^{(2)}=&\int dm \,\mathcal{A}_m\int dx^4e^{3Ht}\left[\dot{\bar{h}}^{(m)ij}\,\dot{\bar{h}}^{(m)}_{ij}\notag\right.\\
&\left.-e^{-2Ht}\partial^k\bar{h}^{(m)ij}\,\partial_k\bar{h}^{(m)}_{ij}-m^2\bar{h}^{(m)ij}\,\bar{h}^{(m)}_{ij}\right],\label{effaction2}
\end{align}
where $\mathcal{A}_m$ is an undetermined coefficient for each mode. On the other hand, the second-order effective action for the matter source, $T_{ij}$, can be obtained from the last term of the action (\ref{action}),
\begin{equation}
S_m^{(2)}=\frac12\int dm\int d^4x\,e^{Ht}\,T_{ij}\bar{h}^{(m)ij}.
\end{equation}
Comparing the effective field equation deduced from the action $S_h^{(2)}+S_m^{(2)}$ with the one given in Eq.(\ref{4Deffequation}), we are able to fix the coefficient $\mathcal{A}_m$ in terms of the normalization constant $\chi_m$, which reads
\begin{equation}
\mathcal{A}_m=\frac{(1-\beta)}{8\kappa^2_5}\,\mathcal{E}^{-2}_m(0)\,\chi_m.\label{coefficient}
\end{equation}
As a result, since the overall coefficient $\mathcal{A}_m$ will compensate for the degrees of freedom of $\chi_m$, the specific choice of $\chi_m$ will not affect the amplitude of the tensor perturbations and therefore those of the gravitational waves. In this paper, we choose $\chi_m=1/(1-\beta)$ in the normalization condition (\ref{sp2}) [see also Eq.(\ref{sp})]. Consequently, using Eq.(\ref{coefficient}), the overall coefficient $\mathcal{A}_m$ of the effective action (\ref{effaction2}) reads
\begin{equation}
\mathcal{A}_m=\frac{\mathcal{E}^{-2}_m(0)}{8\kappa_5^2}.
\end{equation}
The parameter $\chi_m$ has been chosen such that we obtain the same overall normalized factor in the scalar product used in Ref.~\cite{Dufaux:2004qs}.

\end{document}